\documentclass{aa} 
\usepackage{graphicx}
\usepackage{txfonts}
\usepackage[colorlinks=true, linkcolor=blue, citecolor=blue]{hyperref}
\begin{document}

   \title{Increasing the raw contrast of VLT/SPHERE with the dark-hole technique}

   \subtitle{II. On-sky wavefront correction and coherent differential imaging}

   \author{A. Potier\inst{1}\thanks{Based on observations collected at the European Southern Observatory, Chile, 108.22NJ}
          \and
          J. Mazoyer\inst{2}
          \and
          Z. Wahhaj\inst{3}
          \and
          P. Baudoz\inst{2}
          \and
          G. Chauvin\inst{4}
          \and
          R. Galicher\inst{2}
          \and
          G. Ruane\inst{1}
          }

   \institute{Jet Propulsion Laboratory, California Institute of Technology, 4800 Oak Grove Drive, Pasadena, CA 91109\\
              \email{axel.q.potier@jpl.nasa.gov}
         \and
           LESIA, Observatoire de Paris, Université PSL, CNRS, Université Paris Cité, Sorbonne Université, 5 place Jules Janssen, 92195 Meudon, France
        \and
            European Southern Observatory, Alonso de Cordova 3107, Vitacura, Santiago, Chile
        \and
            Laboratoire Lagrange, Université Cote d’Azur, CNRS, Observatoire de la Cote d’Azur, 06304 Nice, France}

   \date{Received XXX; accepted XXX}

\abstract
  {Direct imaging of exoplanets takes advantage of state-of-the-art adaptive optics (AO) systems, coronagraphy, and post-processing techniques. Coronagraphs attenuate starlight to mitigate the unfavorable flux ratio between an exoplanet and its host star. AO systems provide diffraction-limited images of point sources and minimize optical aberrations that would cause starlight to leak through coronagraphs. Post-processing techniques then estimate and remove residual stellar speckles due to hardware limitations, such as noncommon path aberrations (NCPAs) and diffraction from telescope obscurations, and identify potential companions.}
   {We aim to demonstrate an efficient method to minimize the speckle intensity due to NCPAs and the underlying stellar diffraction pattern during an observing night on the Spectro-Polarimetric High-contrast Expolanet REsearch (SPHERE) instrument at the Very Large Telescope (VLT) instrument without any hardware modifications.}
   {We implement an iterative dark-hole (DH) algorithm to remove stellar speckles on-sky before a science observation. It uses a pair-wise probing estimator and a controller based on electric field conjugation, originally developed for space-based application. This work presents the first such on-sky minimization of speckles with a DH technique on SPHERE.}
   {We show the standard deviation of the normalized intensity in the raw images is reduced by a factor of up to five in the corrected region with respect to the current calibration strategy under median conditions for VLT. This level of contrast performance obtained with only 1 min of exposure time reaches median performances on SPHERE that use post-processing methods requiring $\sim$1h-long sequences of observations. The resulting raw contrast improvement provides access to potentially fainter and lower-mass exoplanets closer to their host stars. We also present an alternative a posteriori calibration method that takes advantage of the starlight coherence and improves the post-processed contrast levels rms by a factor of about three with respect to the raw images.}
   {This on-sky demonstration represents a decisive milestone for the future design, development, and observing strategy of the next generation of ground-based exoplanet imagers for 10-m to 40-m telescopes.}


   \maketitle
%\blfootnote{\textcopyright 2022. All rights reserved.}
%
%-------------------------------------------------------------------

\section{Introduction}
The imaging of faint stellar companions, exoplanets, and circumstellar disks with ground-based high-contrast imaging instruments, such as the Spectro-Polarimetric High-contrast Expolanet REsearch (SPHERE) instrument at the Very Large Telescope (VLT) \citep{Beuzit2019} and the Gemini Planet Imager (GPI) \citep{Macintosh2015a}, is currently limited by wavefront aberrations and diffraction effects. Extreme adaptive optics (XAO) systems control dynamic optical perturbations caused by atmospheric turbulence and compensate for static phase aberrations introduced by manufacturing defects and misalignments of optical components from the telescope primary mirror to the XAO wavefront sensor (WFS). Classical AO systems split the beam and analyze the wavefront aberrations in a dedicated optical channel. This two-channel configuration (one for the measurement of the aberrations and one for the science detector) cannot constrain the aberrations introduced by the optical components in the science channel. These noncommon path aberrations \citep[NCPAs,][]{Fusco2006} induce additional starlight leakage that interferes with the underlying stellar diffraction pattern and takes the form of static and quasi-static speckles on the science detector, currently preventing the detection of sources fainter than one millionth of their host star's flux.

The mitigation of static speckles caused by NCPAs has been extensively studied by the community. First, postprocessing techniques such as angular \citep[ADI,][]{Marois2006}, spectral \citep[SDI,][]{Racine1999}, polarimetric \citep[PDI,][]{Kuhn2001}, and reference-star \citep[RDI,][]{Lafreniere2009} differential imaging are broadly employed to enhance detection levels in the images. Both ADI and SDI, routinely used for the detection of exoplanets, suffer from self-subtraction of the off-axis sources at small angular separations \citep{Esposito2014}. ADI requires additional observing time to increase the total parallactic angle rotation, limiting the total observing efficiency. To enhance the contrast in the images before post-processing, the NCPAs are also currently compensated, prior to the observing night, with phase retrieval techniques \citep{Gonsalves1982, Paxman1992, Sauvage2007}. Methods to compensate for the NCPAs with the XAO high-order deformable mirror (HODM) in parallel with the science observations is an active field of research. Science cameras running at high rates allow for techniques that compare WFS telemetry with focal plane images to characterize the optical system and measure the NCPAs \citep{Rodack2021, Skaf2022}. But, the long read-out time of science detectors prevents the use of such methods with most of the high-contrast imagers. Other techniques based on focal plane WFSs, initially developed for space-based applications, have also been tested in dynamic conditions and implemented on different ground-based instruments. For instance, the self-coherent camera \citep[SCC,][]{Baudoz2006} has been validated on-sky to correct for low-order aberrations \citep{Galicher2019}. Higher-order corrections have been demonstrated on-sky using speckle nulling techniques \citep{Borde2006, Martinache2014, Bottom2016}, as well as the Zernike sensor for Extremely Low-level Differential Aberration \citep[ZELDA, ][]{NDiaye2016, Vigan2019}. The coronagraphic phase diversity (COFFEE), as well as pair-wise probing (PWP), associated with the electric field conjugation (EFC) algorithm \citep{GiveOn2007SPIE} were previously only tested in the laboratory with a rotating phase plate simulating turbulence residuals \citep{Potier2019, Herscovici2019} and to calibrate state-of-the-art instruments using their internal sources \citep{Paul2014, Matthews2017, Potier2020b}.

The combination of PWP with EFC aims to minimize the speckle intensity in parts of the science detector called dark holes (DHs). PWP estimates the complex focal-plane electric field (E-field) through a temporal modulation of the intensity by introducing phase-shifted surface shapes to the HODM, resulting in an interferometric measurement in the image plane. Once estimated, the static E-field is minimized by the EFC algorithm, which determines the HODM settings that destructively interfere with the measured E-field in the DH region. As shown in \cite{Potier2020b}, the technique applied to half of the field of view (commonly called a half DH) enables the compensation of both the static phase and amplitude aberrations, as well as the underlying stellar diffraction pattern, thus improving the contrast in the raw images (i.e., the intensity of the stellar residuals normalized by the off-axis point spread function of the same star). While this method has been shown to be suitable for correcting aberrations with a stable wavefront at extremely high-contrast ratios \citep{Trauger2007, Seo2019, Potier2020a}, its application is more challenging during on-sky observations of stars with ground-based telescopes due to the varying conditions between the different images required by PWP. This is likely why the loop of the iterative PWP+EFC technique has never been successfully closed on-sky previously. However, we recently demonstrated in the lab that leaving the dynamic speckles to average in a halo through long-exposure imaging should overcome previous limitations and enable the correction of speckles that evolve slower than the servo-loop speed \citep{Singh2019}.

In this paper, we describe the on-sky results obtained with SPHERE where we successfully minimized the speckle intensity with PWP+EFC. In Sec.~\ref{sec:onsky_correction}, we describe the direct on-sky compensation of the NCPAs in closed loop. In Sec.~\ref{sec:CDI}, we demonstrate that the PWP estimations can be used to implement an alternative coherent differential imaging (CDI) post-processing method that improves the exoplanet detection limit in the images, even on the uncorrected parts of the science detector.
%--------------------------------------------------------------------
\section{On-sky closed loop of the DH algorithm}
\label{sec:onsky_correction}
\subsection{Method}
\label{subsec:Method_onsky}
The observations were performed during the second half of the night of February 15, 2022 starting at 5~a.m. UTC. The seeing ranged from 0.56" to 1.2" at 500~nm according to the Paranal DIMM-Seeing monitor, slowly degrading over time with an average of $\sim$0.8". This value is typical of median conditions at Paranal. The coherence time remained stable, varying between 8 and 10\,ms during the full observing sequence. Wind originated from between the north-east and the north-west, with a speed below 5~m.s$^{-1}$ throughout the observations. These good conditions allowed us to use the small pinhole for spatial filtering in front on the Shack-Hartmann (SH) WFS \citep{Poyneer2004}. We observed HIP~57013 (\mbox{RA=11 41 19.79}, \mbox{$\delta$= -43 05 44.40}, R=5.5, H=5.5), a bright early-type A0V member of the Sco-Cen association and part of the BEAST survey \citep{Janson2021}, using the H3 filter ($\lambda_0 = 1667$~nm, $\Delta\lambda = 54$~nm). We attenuated the starlight with an Apodized Pupil Lyot Coronagraph \citep[APLC,][]{Soummer2005}, whose focal plane occultor is 185~mas in diameter.

Starting from the original SPHERE wavefront calibration based on phase retrieval, we ran the PWP+EFC focal plane wavefront sensing and control algorithms, following the method described by \cite{Potier2020b}. The pair-wise probes and EFC solutions were applied on the HODM by changing the SH WFS reference slopes while the AO system ran in closed loop. The pair-wise probes consisted in the vertical displacement of one individual actuator located within the Lyot stop opening, and poked with a peak-to-valley amplitude of $\pm$400\,nm. We used two probes per iteration, which were distributed vertically to create a top DH as described in \cite{Potier2020b}. We used an exposure time of 64\,s, which averages the intensity variations due to atmospheric turbulence. Assuming the atmospheric turbulence was in steady-state during one iteration and small AO residuals with respect to the wavelength \citep{Singh2019}, the turbulent halo was ignored by PWP when the difference between each pair of diversity images was calculated.

\begin{figure*}[!ht]
    \centering
    \includegraphics[width=\linewidth]{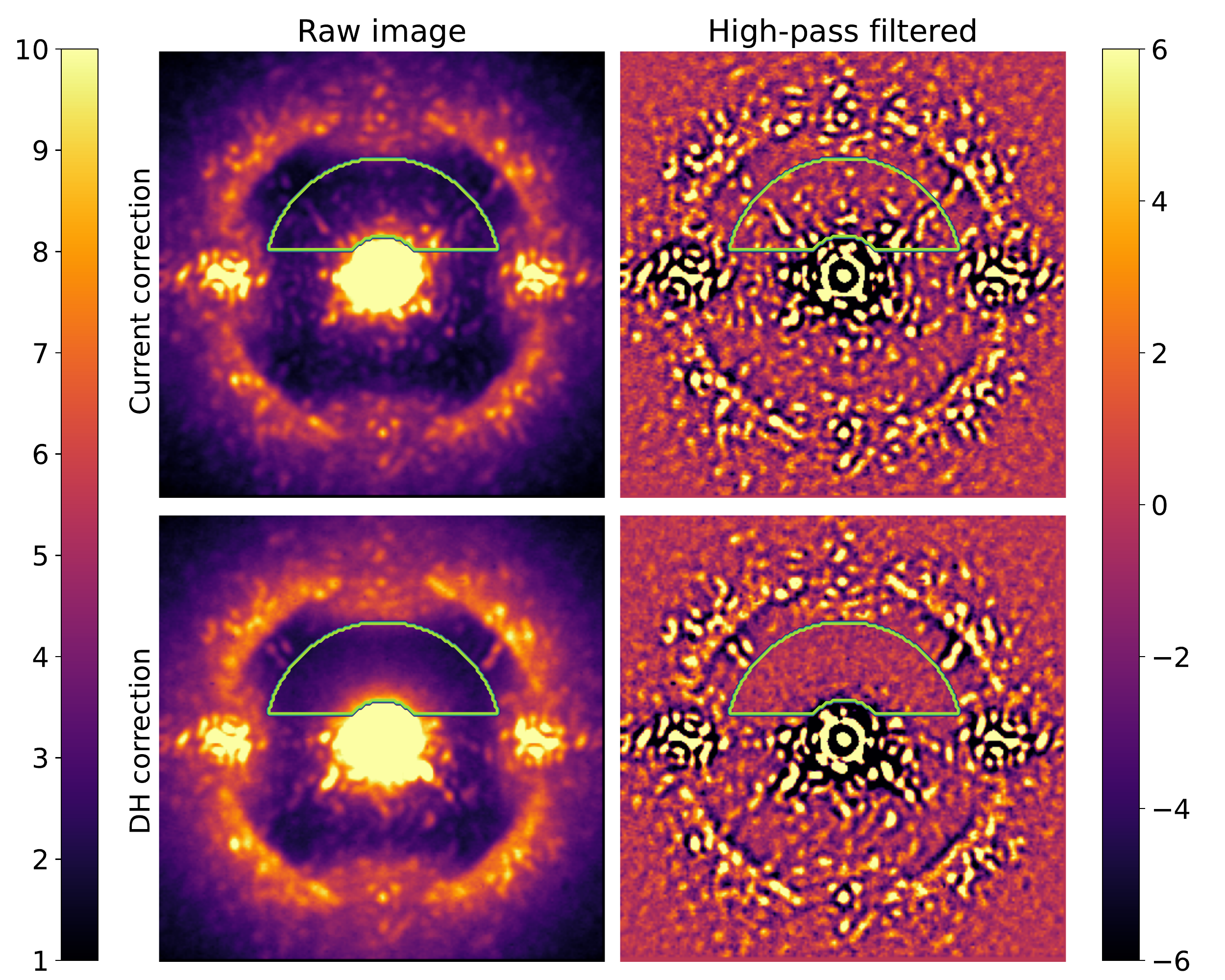}
    \caption{Normalized intensity images in linear scale obtained before (top row) and after (bottom row) the on-sky DH correction. (Left column) Raw image with bad pixels removed. (Right column) Same as for the left column, but high-pass filtered using a Gaussian kernel with a standard deviation of 0.57$\lambda_0/D$. The images are normalized by the maximum of the off-axis PSF, and multiplied by $10^5$ and $10^6$ for the left and right images, respectively.  The corrected region is encircled in green.}
    \label{fig:OnSkyDH}%
\end{figure*}

The corrected region (DH) is an annulus with an inner and outer edge of 5~$\lambda_0/D$ and 15~$\lambda_0/D$, respectively, which corresponds to angular separations ranging from 220~mas to 650~mas. This choice aims to minimize the risk of the loop being perturbed by the starlight leakage due to AO aliasing \citep[Fig.~6 of][]{Cantalloube2019} or by the wind driven halo \citep{Cantalloube2020}. At each iteration, the EFC solution is filtered through a Tikhonov regularization \citep{Pueyo2009}, which suppresses modes higher than the 400th mode out of 1377 total modes controlled by SPHERE. A low servo-loop integral gain of 0.1 was used to account for biases in the PWP, especially those induced by small unexpected fluctuations in the intensity (mainly originating from the wind driven halo) during the acquisition of the probe images and errors in the model of light propagation. Given the apparent stability of the loop, we increased the gain to 0.2 from the third iteration to expedite the correction. Each probe image acquisition required 80\,s (64\,s exposure and $\sim$16\,s overhead, including readout time) for a total of $\sim$400\,s per iteration (four images required for PWP and one image to measure current raw contrast in an unprobed image). The total time required for the on-sky NCPA calibration could be reduced in the future, either by increasing the servo-loop gain, by decreasing the exposure time for the probed images, or by skipping the unprobed image that we used to verify loop convergence. This optimization will be performed before the official commissioning of this technique on SPHERE.

\subsection{Results}
\label{subsec:Results_onsky}
Figure~\ref{fig:OnSkyDH} shows the on-sky coronagraphic images (left) using the current SPHERE calibration procedure (top) and the closed-loop~DH correction inside the green line after five iterations (bottom). Images are normalized by the maximum of the off-axis point spread function (PSF) measured on the same star (i.e., the no-coronagraph~PSF) to study the results in the ``normalized intensity'' dimensionless metric. Normalized intensity differs from raw contrast in that it does not account for potential spatial variations in the off-axis PSF transmission and morphology. The quasi-static speckles that set the detection limit for current SPHERE observations are effectively removed from the ``DH correction'' image, leaving only the fainter, smooth AO halo. The final raw image is similar to the simulated bottom left image of Fig.~4 in \cite{Potier2020b} where the halo of the SPHERE AO system (SAXO) limits the overall performance after compensating for the static speckles in the DH. The raw image in Fig.~\ref{fig:OnSkyDH} is further limited by the fast-evolving low-order aberrations, whose level was underestimated in the former simulation. To reduce the effect of this halo, and highlight the impact of static and quasi-static speckles in the SPHERE performance, the images are high-pass filtered using a Gaussian kernel with a standard deviation of 2 pixels (0.57$\lambda_0/D$, right column of Fig.~\ref{fig:OnSkyDH}). In the DH, the visible static speckles have been corrected, minimizing the estimated mean coherent intensity from $\sim6\cdot10^{-6}$ to $\sim1\cdot10^{-6}$. The remaining photo-electrons in the smooth regions in the high-pass-filtered DH after an exposure time of 64~s with the science detector mostly come from the photon noise induced by the turbulent halo.

\begin{figure}
    \centering
    \includegraphics[width=\linewidth]{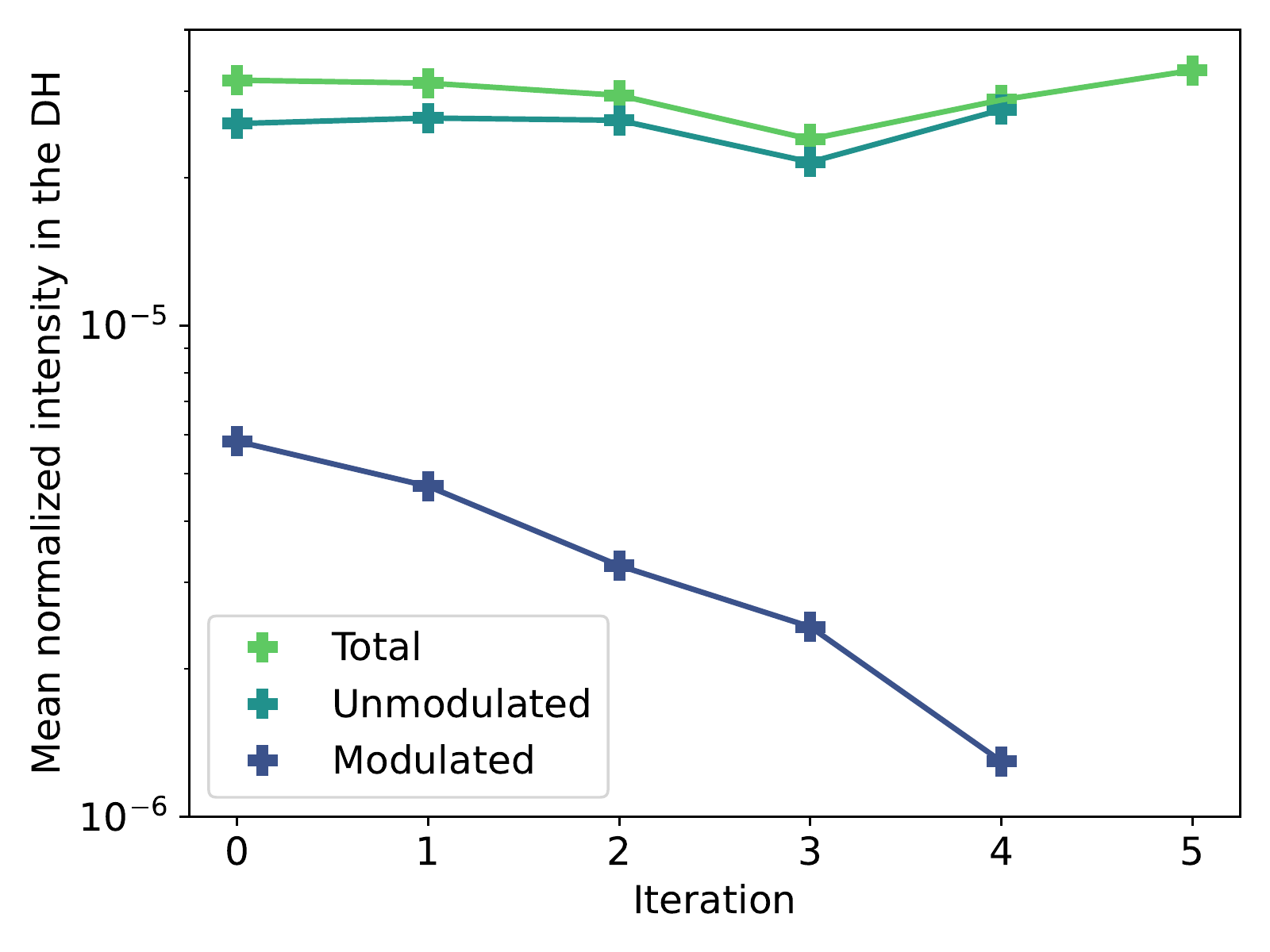}
    \caption{Mean normalized intensity of the total, modulated and unmodulated signals in the DH after each iteration of PWP+EFC and calculated on the unfiltered data. Iteration 0 corresponds to the initial state. PWP has not been applied at iteration 5, preventing the calculation of both the coherent and incoherent components.}
    \label{fig:Contrast_Iteration}%
\end{figure}

We show the mean modulated and unmodulated intensities in the DH at each iteration in Fig.~\ref{fig:Contrast_Iteration}. The modulated component is the absolute value of the E-field squared estimated by PWP, while the unmodulated component represents the difference between the total recorded intensity and the modulated component. The unmodulated component therefore includes: 1)~astrophysical sources located in the field of view; 2)~calibration errors in PWP, such as instrument model inaccuracies; 3)~the field of speckles whose life time is smaller than the measurement rate of PWP ($\sim$5~min), such as the turbulent halo; and 4)~chromatic residuals. Figure~\ref{fig:Contrast_Iteration} demonstrates the convergence of the PWP+EFC closed-loop algorithm because the averaged modulated component in the DH is minimized. However, it also shows that the averaged level of starlight residual in the total image remains almost constant at $\sim3\cdot10^{-5}$ throughout the process. Indeed, even though the processed high-contrast imaging data are limited by the effects of static and quasi-static aberrations at small separations, the smooth halo of turbulence is often the dominating term in the raw images. This halo is removed in post-processing through the filtering of the low spatial frequency residuals or with differential imaging, leaving behind only its associated photon noise.
\begin{figure}
    \centering
    \includegraphics[width=\linewidth]{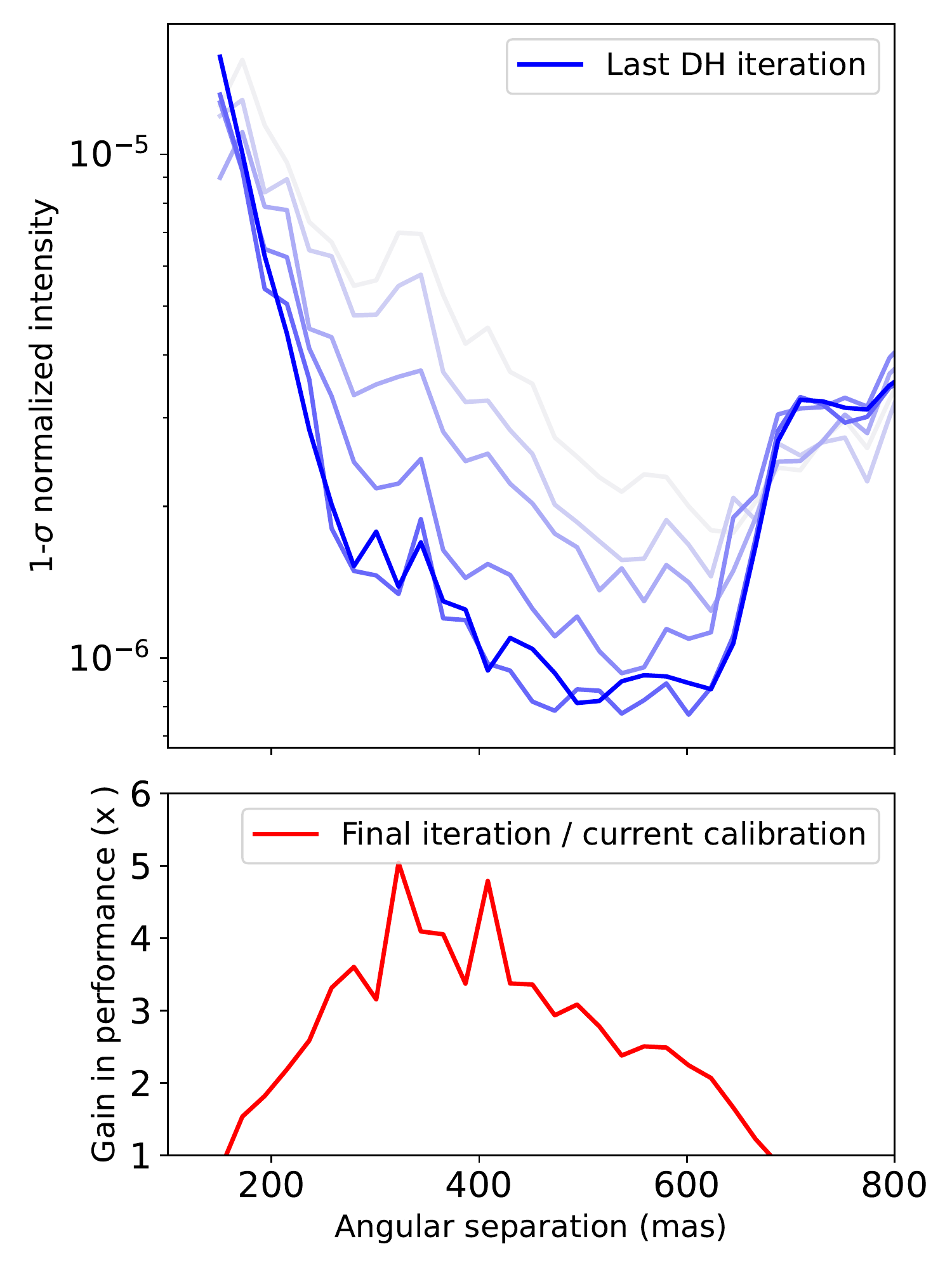}
    \caption{Radial performance of the PWP+EFC closed-loop algorithm in the top DH. Top: 1-$\sigma$ standard deviation of the normalized high-pass-filtered images obtained at each iteration of the PWP+EFC on-sky closed loop in the top DH. Bottom: Gain in performance with respect to the current SPHERE calibration (iteration 0).}
    \label{fig:Contrast_Separation_onsky}%
\end{figure}

We also show the radial profile of the filtered images (calculated as the standard deviation in azimuthal rings of size $\lambda/2D$ in the DH regions versus the angular separation) throughout the different PWP+EFC iterations in Fig.~\ref{fig:Contrast_Separation_onsky}. Detection performance in the DH region has been improved from a factor of two to a factor of five in that area. We obtain the best improvement factor (more than three) between 245 and 500~mas. The normalized intensity reaches a level below $10^{-6}$ between 410 and 635~mas. Although we did not try to minimize the starlight below 220~mas, we noticed an improvement between 150 and 220~mas, which we attribute to changing observing conditions (turbulence, wind driven halo, quasi-statics) as iterations progress.

\section{Coherent differential imaging}
\label{sec:CDI}

\subsection{Method}
\label{subsec:Method_CDI}

\begin{figure*}
    \centering
    \includegraphics[width=13.8cm]{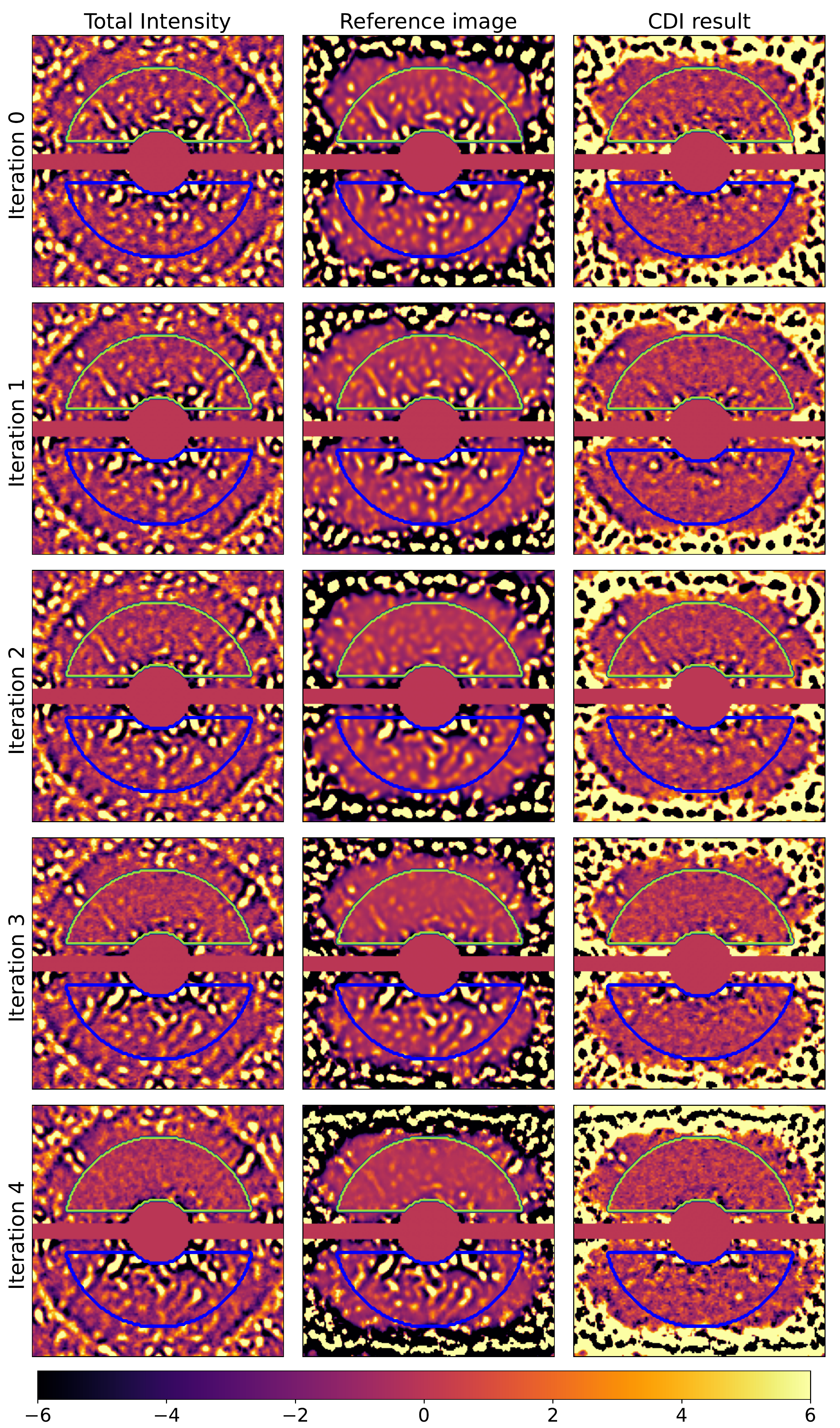}
    \caption{Image decomposition with CDI. Left: Intensity images at each iteration of PWP+EFC. Center: Modulated component measured with PWP at each iteration. Right: CDI results that are calculated as the subtraction between the total intensity and the modulated component (reference image) estimated by PWP. The modulated component was rescaled to minimize the result of CDI in the region encircled in blue. The DH region is encircled in green. All the images were then filtered using a high-pass Gaussian function whose standard deviation is 0.57$\lambda/D$, to remove the halo of turbulence, and multiplied by $10^6$.}
    \label{fig:CDI_steps}%
\end{figure*}

\begin{figure}
    \centering
    \includegraphics[width=\linewidth]{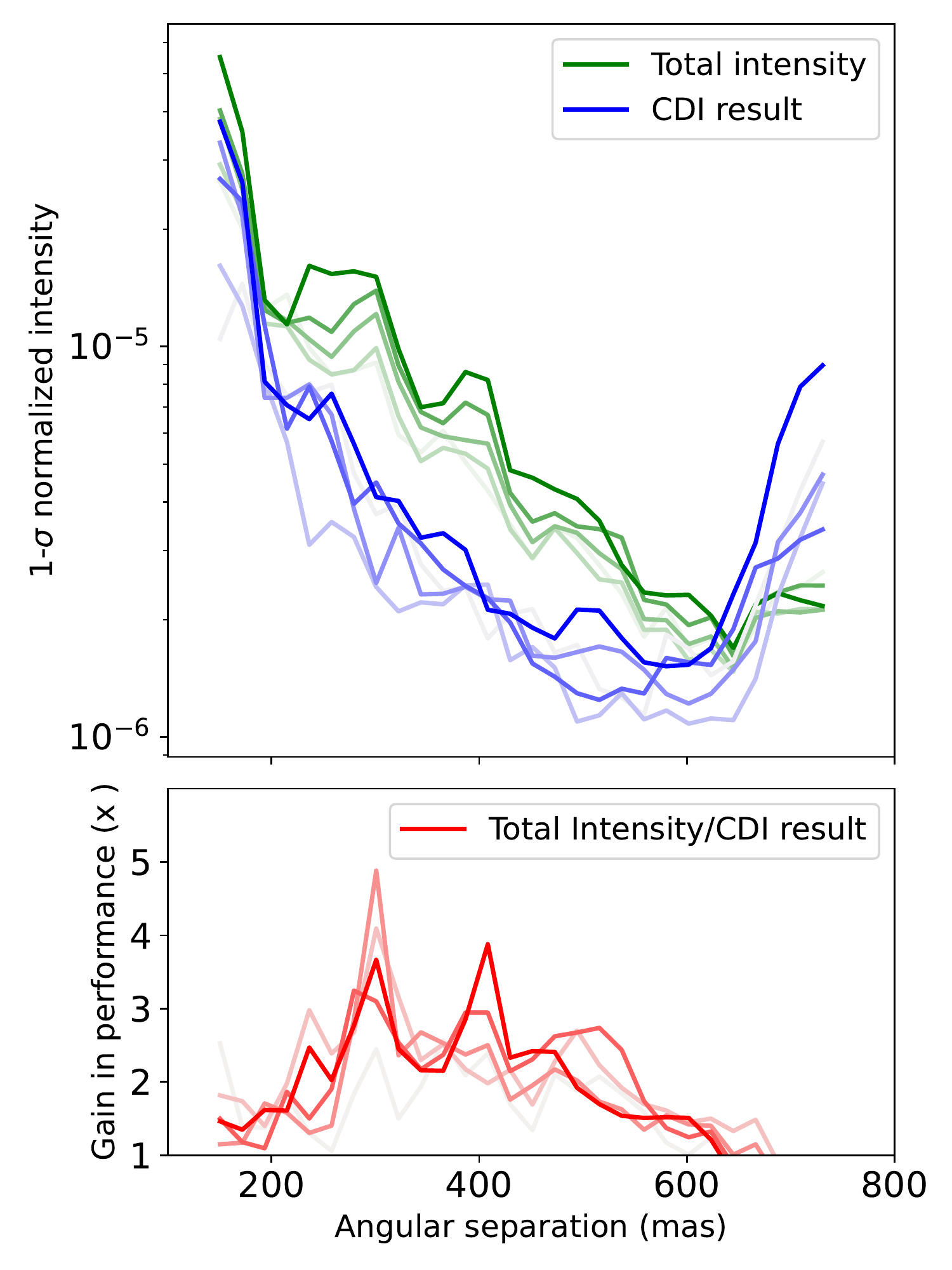}
    \caption{Radial performance of CDI. Top: 1-$\sigma$ standard deviation of the normalized high-pass-filtered total intensity (green) and CDI result (blue) calculated in the blue region in Fig.~\ref{fig:CDI_steps}. The curves are darker along the PWP+EFC iterations described in Sec.~\ref{sec:onsky_correction}. Bottom: Gain in performance of CDI with respect to the total intensity image at each iteration.}
    \label{fig:Contrast_Separation_CDI}%
\end{figure}

PWP can also be used alone to further suppress stellar speckles in the images in post-processing. Indeed, the stellar E-field estimated by PWP can serve as a reference image to be subtracted from the remaining total intensity, thereby improving the S/N of the astrophysical signal that is not coherent with the starlight. This technique called coherent differential imaging (CDI) has been demonstrated with many focal plane wavefront sensors \citep{Baudoz2006, Bottom2017, Jovanovic2018}. CDI can be used to further remove residual speckles inside the DH. In our case, the loop has converged and no more speckles are to be calibrated in post-processing.

In this section, we demonstrate the ability of CDI to enhance the contrast in regions that are not corrected during the PWP+EFC process. We also propose an optimization of the CDI technique to account for the changing atmospheric transmission over time between the recording of the total intensity images and the probe images. This simple algorithm will be investigated further in the near future to quantify potential over-subtraction of the astrophysical signal. At each iteration, the PWP estimation of the focal plane E-field (or reference image) is numerically recentered with respect to the total intensity signal to account for any drift on the science detector during the probing steps. It is then rescaled to minimize the mean intensity of the high-pass-filtered CDI result, in the region symmetrically opposed to the top DH described in Sec.~\ref{sec:onsky_correction}. This area is arbitrarily chosen for the sake of illustration, while any region of the science detector well sensed by PWP could be selected. This optimization assumes that most of the total signal in that area is caused by stellar residuals and is, therefore, coherent. This optimization should not be strongly affected by a planet or other astrophysical object present in the field of view but injection of fake signal in the data is out of the scope of this paper. The estimated mean scaling factor is 1.0, which demonstrates that the calibration of the probes was accurate. The standard deviation of the scaling factor over the iterations is 0.1, which we attribute to varying atmospheric transmissions. The scaling method, reprocessed at each iteration, will be used in the PWP+EFC closed loop in the future for a faster convergence of the DH correction.

\subsection{Results}
\label{subsec:Results_CDI}

The high-pass-filtered total intensity, the optimized reference image, and the CDI result at each iteration are shown in Fig.~\ref{fig:CDI_steps}. Qualitatively, the static speckle intensity in the CDI results is minimized with respect to the total intensity in the region sensed by PWP. The DH geometry is defined either by the regions of the science detector reachable by PWP with the chosen probes, or by the spatial frequencies properly corrected by the AO system. For instance, the CDI performance is limited in the central region along the horizontal axis for the former reason (see Fig.~3 in \citealp{Potier2020b}) and in the regions above the AO cutoff frequency or where aliasing caused by the SH WFS dominates for the latter. Also, the result of CDI is limited by bright residual speckles at small separations. We conclude that PWP is not as sensitive to these residuals, either because they are caused by large wavefront errors that violate the linear approximation made by PWP \citep{GiveOn2007SPIE}, or because they are too close to the fast-evolving low-order aberrations that violate the steady-state assumption required by PWP. For the former case, in principle the iterative PWP+EFC algorithm should be able to correct such aberrations during the observation. Therefore, better performance of the CDI algorithm is expected after a few iterations of DH correction in the targeted region. This approach is well demonstrated by the CDI result at iterations three and four, where the speckles in the top DH are almost perfectly removed.

Figure~\ref{fig:Contrast_Separation_CDI} shows the radial profile of both the total intensity and the result of CDI in the optimized region (encircled in blue in Fig.~\ref{fig:CDI_steps}). The CDI technique enhances the contrast level in the post-processed images between 150 and 640~mas for each estimation of the coherent signal with PWP. With this promising result, we envision observing strategies where PWP is applied at a regular cadence and the resulting CDI results are used as inputs for post-processing methods, or for a direct analysis. However, the contrast after applying CDI degrades with PWP+EFC iterations. Indeed, the minimization of the speckle intensity due to amplitude aberrations in the top DH with one HODM is achieved by degrading the contrast in the bottom part of the field of view. CDI then starts from a higher speckle intensity, with more noise, in the bottom part of the image. Hence, applying CDI before any half DH correction would be more efficient on the ``degrade'' part of the image. The implementation of CDI should, therefore, be defined depending on the science objective and telescope time. Ideally, either CDI would be applied on top of a DH active-wavefront correction for science analysis in that region, requiring the implementation of PWP+EFC successively in different DHs to cover a broader region; or CDI would be directly applied in a larger area by applying PWP alone, without EFC correction, at a regular cadence. The former case would provide the best contrast performance, while the latter would reduce the required telescope time.

\section{Conclusions}

We present two promising implementations of a dark hole~(DH) technique on VLT/SPHERE. First, relying on an efficient adaptive optics~(AO) system combined with good observing conditions, we demonstrated a pair-wise probing ~(PWP) + electric field conjugation~(EFC) algorithm directly on-sky that minimized the static speckle field intensity in a defined region using VLT/SPHERE. The standard deviation of the normalized intensity was improved by a factor of up to five in the DH. Therefore, 1min-exposure images, after the PWP+EFC correction is applied (30min here), achieve the same level of performance as ADI with 1h-sequences of observations. This promising technique is expected to be even more robust with the improvement of the AO systems planned for the next generation of high-contrast imaging instruments \citep{Chilcote2018, Boccaletti2020}. Second, we showed that estimation of the coherent light in the field of view with PWP could be used to further calibrate the static speckles in post-processing through coherent differential imaging~(CDI). PWP could also be used at a regular cadence during a high-contrast observation to estimate the spatial distribution of the coherent speckles and then subtract them from the raw images. The residuals, containing the incoherent signal of interest, could then be further analyzed with conventional methods such as angular~(ADI), spectral~(SDI), or reference~(RDI) differential imaging.

Here, the performance was quantified via a simple high-pass-filtered image. Therefore, the~CDI method also solves the self-subtraction issue that emerges when performing ADI at small angular separations with a coronagraph instrument, such as those that will equip the upcoming extremely large telescopes \citep{Baudoz2014, Fitzgerald2019, Kasper2021}. We will assess this capability in future experiments, where the technique will target the inner edges of known circumstellar disks or face-on disks whose morphologies would be altered by ADI and SDI post-processing techniques. We will also combine this method with star-hopping algorithms \citep{Wahhaj2021} in the near future to understand the speckle lifetime and how the DH correction is impacted by the telescope alignment from one target to another. All of these experiments aim to determine the most optimized strategy of observation for the search for and the characterization of exoplanets in direct imaging.
% in the least amount of time overheads for the astronomers. 

\begin{acknowledgements}
The authors would like to thanks Thierry Fusco, Jean-Fran\c{c}ois Sauvage and Anne Costille for fruitful discussions about SAXO as well as Arthur Vigan and Julien Milli for their insights regarding internal turbulence in SPHERE. AP and GR's research was carried out at the Jet Propulsion Laboratory, California Institute of Technology, under a contract with the National Aeronautics and Space Administration (80NM0018D0004). This work was supported by an ECOS-CONICYT grant (\#C20U02).
\end{acknowledgements}

\bibliographystyle{aa} % style aa.bst
\bibliography{bib_GS} % your references Yourfile.bib

\begin{thebibliography}{44}
\expandafter\ifx\csname natexlab\endcsname\relax\def\natexlab#1{#1}\fi

\bibitem[{{Baudoz} {et~al.}(2006){Baudoz}, {Boccaletti}, {Baudrand}, \&
  {Rouan}}]{Baudoz2006}
{Baudoz}, P., {Boccaletti}, A., {Baudrand}, J., \& {Rouan}, D. 2006, in IAU
  Colloq. 200: Direct Imaging of Exoplanets: Science {\&} Techniques, ed.
  C.~{Aime} \& F.~{Vakili}, 553--558

\bibitem[{{Baudoz} {et~al.}(2014){Baudoz}, {Boccaletti}, {Lacour}, {Galicher},
  {Cl{\'e}net}, {Gratadour}, {Gendron}, {Buey}, {Rousset}, {Hartl}, \&
  {Davies}}]{Baudoz2014}
{Baudoz}, P., {Boccaletti}, A., {Lacour}, S., {et~al.} 2014, Proc. SPIE, 9147,
  91479E

\bibitem[{{Beuzit} {et~al.}(2019){Beuzit}, {Vigan}, {Mouillet}, {Dohlen},
  {Gratton}, {Boccaletti}, {Sauvage}, {Schmid}, {Langlois}, {Petit},
  {Baruffolo}, {Feldt}, {Milli}, {Wahhaj}, {Abe}, {Anselmi}, {Antichi},
  {Barette}, {Baudrand}, {Baudoz}, {Bazzon}, {Bernardi}, {Blanchard}, {Brast},
  {Bruno}, {Buey}, {Carbillet}, {Carle}, {Cascone}, {Chapron}, {Charton},
  {Chauvin}, {Claudi}, {Costille}, {De Caprio}, {de Boer}, {Delboulb{\'e}},
  {Desidera}, {Dominik}, {Downing}, {Dupuis}, {Fabron}, {Fantinel}, {Farisato},
  {Feautrier}, {Fedrigo}, {Fusco}, {Gigan}, {Ginski}, {Girard}, {Giro},
  {Gisler}, {Gluck}, {Gry}, {Henning}, {Hubin}, {Hugot}, {Incorvaia}, {Jaquet},
  {Kasper}, {Lagadec}, {Lagrange}, {Le Coroller}, {Le Mignant}, {Le Ruyet},
  {Lessio}, {Lizon}, {Llored}, {Lundin}, {Madec}, {Magnard}, {Marteaud},
  {Martinez}, {Maurel}, {M{\'e}nard}, {Mesa}, {M{\"o}ller-Nilsson}, {Moulin},
  {Moutou}, {Orign{\'e}}, {Parisot}, {Pavlov}, {Perret}, {Pragt}, {Puget},
  {Rabou}, {Ramos}, {Reess}, {Rigal}, {Rochat}, {Roelfsema}, {Rousset}, {Roux},
  {Saisse}, {Salasnich}, {Santambrogio}, {Scuderi}, {Segransan}, {Sevin},
  {Siebenmorgen}, {Soenke}, {Stadler}, {Suarez}, {Tiph{\`e}ne}, {Turatto},
  {Udry}, {Vakili}, {Waters}, {Weber}, {Wildi}, {Zins}, \&
  {Zurlo}}]{Beuzit2019}
{Beuzit}, J.~L., {Vigan}, A., {Mouillet}, D., {et~al.} 2019, Astron.\
  Astrophys., 631, A155

\bibitem[{{Boccaletti} {et~al.}(2020){Boccaletti}, {Chauvin}, {Mouillet},
  {Absil}, {Allard}, {Antoniucci}, {Augereau}, {Barge}, {Baruffolo}, {Baudino},
  {Baudoz}, {Beaulieu}, {Benisty}, {Beuzit}, {Bianco}, {Biller}, {Bonavita},
  {Bonnefoy}, {Bos}, {Bouret}, {Brandner}, {Buchschache}, {Carry},
  {Cantalloube}, {Cascone}, {Carlotti}, {Charnay}, {Chiavassa}, {Choquet},
  {Clenet}, {Crida}, {De Boer}, {De Caprio}, {Desidera}, {Desert}, {Delisle},
  {Delorme}, {Dohlen}, {Doelman}, {Dominik}, {Orazi}, {Dougados}, {Doute},
  {Fedele}, {Feldt}, {Ferreira}, {Fontanive}, {Fusco}, {Galicher}, {Garufi},
  {Gendron}, {Ghedina}, {Ginski}, {Gonzalez}, {Gratadour}, {Gratton},
  {Guillot}, {Haffert}, {Hagelberg}, {Henning}, {Huby}, {Janson}, {Kamp},
  {Keller}, {Kenworthy}, {Kervella}, {Kral}, {Kuhn}, {Lagadec}, {Laibe},
  {Langlois}, {Lagrange}, {Launhardt}, {Leboulleux}, {Le Coroller}, {Li Causi},
  {Loupias}, {Maire}, {Marleau}, {Martinache}, {Martinez}, {Mary}, {Mattioli},
  {Mazoyer}, {Meheut}, {Menard}, {Mesa}, {Meunier}, {Miguel}, {Milli}, {Min},
  {Molliere}, {Mordasini}, {Moretto}, {Mugnier}, {Muro Arena}, {Nardetto},
  {Diaye}, {Nesvadba}, {Pedichini}, {Pinilla}, {Por}, {Potier}, {Quanz},
  {Rameau}, {Roelfsema}, {Rouan}, {Rigliaco}, {Salasnich}, {Samland},
  {Sauvage}, {Schmid}, {Segransan}, {Snellen}, {Snik}, {Soulez}, {Stadler},
  {Stam}, {Tallon}, {Thebault}, {Thiebaut}, {Tschudi}, {Udry}, {van Holstein},
  {Vernazza}, {Vidal}, {Vigan}, {Waters}, {Wildi}, {Willson}, {Zanutta},
  {Zavagno}, \& {Zurlo}}]{Boccaletti2020}
{Boccaletti}, A., {Chauvin}, G., {Mouillet}, D., {et~al.} 2020, arXiv e-prints,
  arXiv:2003.05714

\bibitem[{{Bord{\'e}} \& {Traub}(2006)}]{Borde2006}
{Bord{\'e}}, P.~J. \& {Traub}, W.~A. 2006, Astrophys.\ J., 638, 488

\bibitem[{{Bottom} {et~al.}(2016){Bottom}, {Femenia}, {Huby}, {Mawet},
  {Dekany}, {Milburn}, \& {Serabyn}}]{Bottom2016}
{Bottom}, M., {Femenia}, B., {Huby}, E., {et~al.} 2016, Proc. SPIE, 9909,
  990955

\bibitem[{{Bottom} {et~al.}(2017){Bottom}, {Wallace}, {Bartos}, {Shelton}, \&
  {Serabyn}}]{Bottom2017}
{Bottom}, M., {Wallace}, J.~K., {Bartos}, R.~D., {Shelton}, J.~C., \&
  {Serabyn}, E. 2017, \mnras, 464, 2937

\bibitem[{{Cantalloube} {et~al.}(2019){Cantalloube}, {Dohlen}, {Milli},
  {Brandner}, \& {Vigan}}]{Cantalloube2019}
{Cantalloube}, F., {Dohlen}, K., {Milli}, J., {Brandner}, W., \& {Vigan}, A.
  2019, The Messenger, 176, 25

\bibitem[{{Cantalloube} {et~al.}(2020){Cantalloube}, {Farley}, {Milli},
  {Bharmal}, {Brandner}, {Correia}, {Dohlen}, {Henning}, {Osborn}, {Por},
  {Su{\'a}rez Valles}, \& {Vigan}}]{Cantalloube2020}
{Cantalloube}, F., {Farley}, O.~J.~D., {Milli}, J., {et~al.} 2020, \aap, 638,
  A98

\bibitem[{{Chilcote} {et~al.}(2018){Chilcote}, {Bailey}, {De Rosa},
  {Macintosh}, {Nielsen}, {Norton}, {Millar-Blanchaer}, {Graham}, {Marois},
  {Pueyo}, {Rameau}, {Savransky}, \& {Veran}}]{Chilcote2018}
{Chilcote}, J.~K., {Bailey}, V.~P., {De Rosa}, R., {et~al.} 2018, Proc. SPIE,
  10702, 1070244

\bibitem[{{Esposito} {et~al.}(2014){Esposito}, {Fitzgerald}, {Graham}, \&
  {Kalas}}]{Esposito2014}
{Esposito}, T.~M., {Fitzgerald}, M.~P., {Graham}, J.~R., \& {Kalas}, P. 2014,
  \apj, 780, 25

\bibitem[{{Fitzgerald} {et~al.}(2019){Fitzgerald}, {Bailey}, {Baranec},
  {Batalha}, {Benneke}, {Beichman}, {Brandt}, {Chilcote}, {Chun}, {Crossfield},
  {Currie}, {Davis}, {Dekany}, {Delorme}, {Dong}, {Doyon}, {Dressing},
  {Echeverri}, {Fortney}, {Frazin}, {Guyon}, {Hashimoto}, {Hillenbrand},
  {Hinz}, {Howard}, {Jensen-Clem}, {Jovanovic}, {Kawahara}, {Knutson},
  {Konopacky}, {Kotani}, {Lafreni{\`e}re}, {Liu}, {Lozi}, {Lu}, {Males},
  {Marley}, {Marois}, {Mawet}, {Mazin}, {Millar-Blanchaer}, {Mondal},
  {Murakami}, {Murray-Clay}, {Narita}, {Pezzato}, {Pyo}, {Roberts}, {Ruane},
  {Sallum}, {Serabyn}, {Shields}, {Simard}, {Skemer}, {Stelter}, {Tamura},
  {Troy}, {Vasisht}, {Wallace}, {Wang}, {Wang}, \& {Wright}}]{Fitzgerald2019}
{Fitzgerald}, M., {Bailey}, V., {Baranec}, C., {et~al.} 2019, in Bulletin of
  the American Astronomical Society, Vol.~51, 251

\bibitem[{{Fusco} {et~al.}(2006){Fusco}, {Rousset}, {Sauvage}, {Petit},
  {Beuzit}, {Dohlen}, {Mouillet}, {Charton}, {Nicolle}, {Kasper}, {Baudoz}, \&
  {Puget}}]{Fusco2006}
{Fusco}, T., {Rousset}, G., {Sauvage}, J.~F., {et~al.} 2006, Optics Express,
  14, 7515

\bibitem[{{Galicher} {et~al.}(2019){Galicher}, {Baudoz}, {Delorme}, {Mawet},
  {Bottom}, {Wallace}, {Serabyn}, \& {Shelton}}]{Galicher2019}
{Galicher}, R., {Baudoz}, P., {Delorme}, J.~R., {et~al.} 2019, Astron.\
  Astrophys., 631, A143

\bibitem[{{Give'On} {et~al.}(2007){Give'On}, {Kern}, {Shaklan}, {Moody}, \&
  {Pueyo}}]{GiveOn2007SPIE}
{Give'On}, A., {Kern}, B., {Shaklan}, S., {Moody}, D.~C., \& {Pueyo}, L. 2007,
  Proc. SPIE, 6691, 66910A

\bibitem[{Gonsalves(1982)}]{Gonsalves1982}
Gonsalves, R.~A. 1982, Opt. Eng., 21, 829

\bibitem[{{Herscovici-Schiller} {et~al.}(2019){Herscovici-Schiller}, {Sauvage},
  {Mugnier}, {Dohlen}, \& {Vigan}}]{Herscovici2019}
{Herscovici-Schiller}, O., {Sauvage}, J.-F., {Mugnier}, L.~M., {Dohlen}, K., \&
  {Vigan}, A. 2019, Mon.\ Not.\ R.\ Astr.\ Soc., 488, 4307

\bibitem[{{Janson} {et~al.}(2021){Janson}, {Squicciarini}, {Delorme},
  {Gratton}, {Bonnefoy}, {Reffert}, {Mamajek}, {Eriksson}, {Vigan}, {Langlois},
  {Engler}, {Chauvin}, {Desidera}, {Mayer}, {Marleau}, {Bohn}, {Samland},
  {Meyer}, {d'Orazi}, {Henning}, {Quanz}, {Kenworthy}, \&
  {Carson}}]{Janson2021}
{Janson}, M., {Squicciarini}, V., {Delorme}, P., {et~al.} 2021, \aap, 646, A164

\bibitem[{{Jovanovic} {et~al.}(2018){Jovanovic}, {Absil}, {Baudoz}, {Beaulieu},
  {Bottom}, {Cady}, {Carlomagno}, {Carlotti}, {Doelman}, {Fogarty}, {Galicher},
  {Guyon}, {Haffert}, {Huby}, {Jewell}, {Keller}, {Kenworthy}, {Knight},
  {K{\"u}hn}, {Miller}, {Mazoyer}, {N'Diaye}, {Por}, {Pueyo}, {Riggs}, {Ruane},
  {Sirbu}, {Snik}, {Wallace}, {Wilby}, \& {Ygouf}}]{Jovanovic2018}
{Jovanovic}, N., {Absil}, O., {Baudoz}, P., {et~al.} 2018, Proc. SPIE, 10703,
  107031U

\bibitem[{{Kasper} {et~al.}(2021){Kasper}, {Cerpa Urra}, {Pathak}, {Bonse},
  {Nousiainen}, {Engler}, {Heritier}, {Kammerer}, {Leveratto}, {Rajani},
  {Bristow}, {Le Louarn}, {Madec}, {Str{\"o}bele}, {Verinaud}, {Glauser},
  {Quanz}, {Helin}, {Keller}, {Snik}, {Boccaletti}, {Chauvin}, {Mouillet},
  {Kulcs{\'a}r}, \& {Raynaud}}]{Kasper2021}
{Kasper}, M., {Cerpa Urra}, N., {Pathak}, P., {et~al.} 2021, The Messenger,
  182, 38

\bibitem[{{Kuhn} {et~al.}(2001){Kuhn}, {Potter}, \& {Parise}}]{Kuhn2001}
{Kuhn}, J.~R., {Potter}, D., \& {Parise}, B. 2001, Astrophys.\ J.\ Lett., 553,
  L189

\bibitem[{Lafreni{\`{e}}re {et~al.}(2009)Lafreni{\`{e}}re, Marois, Doyon, \&
  Barman}]{Lafreniere2009}
Lafreni{\`{e}}re, D., Marois, C., Doyon, R., \& Barman, T. 2009, Astrophys.\
  J., 694, L148

\bibitem[{{Macintosh} {et~al.}(2014){Macintosh}, {Graham}, {Ingraham},
  {Konopacky}, {Marois}, {Perrin}, {Poyneer}, {Bauman}, {Barman}, {Burrows},
  {Cardwell}, {Chilcote}, {De Rosa}, {Dillon}, {Doyon}, {Dunn}, {Erikson},
  {Fitzgerald}, {Gavel}, {Goodsell}, {Hartung}, {Hibon}, {Kalas}, {Larkin},
  {Maire}, {Marchis}, {Marley}, {McBride}, {Millar-Blanchaer}, {Morzinski},
  {Norton}, {Oppenheimer}, {Palmer}, {Patience}, {Pueyo}, {Rantakyro},
  {Sadakuni}, {Saddlemyer}, {Savransky}, {Serio}, {Soummer},
  {Sivaramakrishnan}, {Song}, {Thomas}, {Wallace}, {Wiktorowicz}, \&
  {Wolff}}]{Macintosh2015a}
{Macintosh}, B., {Graham}, J.~R., {Ingraham}, P., {et~al.} 2014, Proceedings of
  the National Academy of Science, 111, 12661

\bibitem[{{Marois} {et~al.}(2006){Marois}, {Lafreni{\`e}re}, {Doyon},
  {Macintosh}, \& {Nadeau}}]{Marois2006}
{Marois}, C., {Lafreni{\`e}re}, D., {Doyon}, R., {Macintosh}, B., \& {Nadeau},
  D. 2006, Astrophys.\ J., 641, 556

\bibitem[{{Martinache} {et~al.}(2014){Martinache}, {Guyon}, {Jovanovic},
  {Clergeon}, {Singh}, {Kudo}, {Currie}, {Thalmann}, {McElwain}, \&
  {Tamura}}]{Martinache2014}
{Martinache}, F., {Guyon}, O., {Jovanovic}, N., {et~al.} 2014, Pub.\ Astron.\
  Soc.\ Pacific, 126, 565

\bibitem[{{Matthews} {et~al.}(2017){Matthews}, {Crepp}, {Vasisht}, \&
  {Cady}}]{Matthews2017}
{Matthews}, C.~T., {Crepp}, J.~R., {Vasisht}, G., \& {Cady}, E. 2017, Journal
  of Astronomical Telescopes, Instruments, and Systems, 3, 045001

\bibitem[{{N'Diaye} {et~al.}(2016){N'Diaye}, {Soummer}, {Pueyo}, {Carlotti},
  {Stark}, \& {Perrin}}]{NDiaye2016}
{N'Diaye}, M., {Soummer}, R., {Pueyo}, L., {et~al.} 2016, Astrophys.\ J., 818,
  163

\bibitem[{{Paul} {et~al.}(2014){Paul}, {Sauvage}, {Mugnier}, {Dohlen}, {Petit},
  {Fusco}, {Mouillet}, {Beuzit}, \& {Ferrari}}]{Paul2014}
{Paul}, B., {Sauvage}, J.~F., {Mugnier}, L.~M., {et~al.} 2014, Astron.\
  Astrophys., 572, A32

\bibitem[{{Paxman} {et~al.}(1992){Paxman}, {Schulz}, \& {Fienup}}]{Paxman1992}
{Paxman}, R.~G., {Schulz}, T.~J., \& {Fienup}, J.~R. 1992, Journal of the
  Optical Society of America A, 9, 1072

\bibitem[{{Potier} {et~al.}(2019){Potier}, {Baudoz}, {Galicher}, {Huby}, \&
  {Singh}}]{Potier2019}
{Potier}, A., {Baudoz}, P., {Galicher}, R., {Huby}, E., \& {Singh}, G. 2019,
  arXiv e-prints, arXiv:1910.09064

\bibitem[{{Potier} {et~al.}(2020{\natexlab{a}}){Potier}, {Baudoz}, {Galicher},
  {Singh}, \& {Boccaletti}}]{Potier2020a}
{Potier}, A., {Baudoz}, P., {Galicher}, R., {Singh}, G., \& {Boccaletti}, A.
  2020{\natexlab{a}}, \aap, 635, A192

\bibitem[{{Potier} {et~al.}(2020{\natexlab{b}}){Potier}, {Galicher}, {Baudoz},
  {Huby}, {Milli}, {Wahhaj}, {Boccaletti}, {Vigan}, {N'Diaye}, \&
  {Sauvage}}]{Potier2020b}
{Potier}, A., {Galicher}, R., {Baudoz}, P., {et~al.} 2020{\natexlab{b}}, \aap,
  638, A117

\bibitem[{{Poyneer} \& {Macintosh}(2004)}]{Poyneer2004}
{Poyneer}, L.~A. \& {Macintosh}, B. 2004, Optics \& Photonics News, 15, 15

\bibitem[{{Pueyo} {et~al.}(2009){Pueyo}, {Kay}, {Kasdin}, {Groff}, {McElwain},
  {Give'on}, \& {Belikov}}]{Pueyo2009}
{Pueyo}, L., {Kay}, J., {Kasdin}, N.~J., {et~al.} 2009, Appl.\ Opt., 48, 6296

\bibitem[{{Racine} {et~al.}(1999){Racine}, {Walker}, {Nadeau}, {Doyon}, \&
  {Marois}}]{Racine1999}
{Racine}, R., {Walker}, G. A.~H., {Nadeau}, D., {Doyon}, R., \& {Marois}, C.
  1999, Pub.\ Astron.\ Soc.\ Pacific, 111, 587

\bibitem[{{Rodack} {et~al.}(2021){Rodack}, {Frazin}, {Males}, \&
  {Guyon}}]{Rodack2021}
{Rodack}, A.~T., {Frazin}, R.~A., {Males}, J.~R., \& {Guyon}, O. 2021, Journal
  of the Optical Society of America A, 38, 1541

\bibitem[{{Sauvage} {et~al.}(2007){Sauvage}, {Fusco}, {Rousset}, \&
  {Petit}}]{Sauvage2007}
{Sauvage}, J.-F., {Fusco}, T., {Rousset}, G., \& {Petit}, C. 2007, Journal of
  the Optical Society of America A, 24, 2334

\bibitem[{{Seo} {et~al.}(2019){Seo}, {Patterson}, {Balasubramanian}, {Crill},
  {Chui}, {Echeverri}, {Kern}, {Marx}, {Moody}, {Mejia Prada}, {Ruane}, {Shi},
  {Shaw}, {Siegler}, {Tang}, {Trauger}, {Wilson}, \& {Zimmer}}]{Seo2019}
{Seo}, B.-J., {Patterson}, K., {Balasubramanian}, K., {et~al.} 2019, Proc.
  SPIE, 11117, 111171V

\bibitem[{{Singh} {et~al.}(2019){Singh}, {Galicher}, {Baudoz}, {Dupuis},
  {Ortiz}, {Potier}, {Thijs}, \& {Huby}}]{Singh2019}
{Singh}, G., {Galicher}, R., {Baudoz}, P., {et~al.} 2019, Astron.\ Astrophys.,
  631, A106

\bibitem[{{Skaf} {et~al.}(2022){Skaf}, {Guyon}, {Gendron}, {Ahn},
  {Bertrou-Cantou}, {Boccaletti}, {Cranney}, {Currie}, {Deo}, {Edwards},
  {Ferreira}, {Gratadour}, {Lozi}, {Norris}, {Sevin}, {Vidal}, \&
  {Vievard}}]{Skaf2022}
{Skaf}, N., {Guyon}, O., {Gendron}, {\'E}., {et~al.} 2022, \aap, 659, A170

\bibitem[{{Soummer}(2005)}]{Soummer2005}
{Soummer}, R. 2005, \apjl, 618, L161

\bibitem[{{Trauger} \& {Traub}(2007)}]{Trauger2007}
{Trauger}, J.~T. \& {Traub}, W.~A. 2007, Nature (London), 446, 771

\bibitem[{{Vigan} {et~al.}(2019){Vigan}, {N'Diaye}, {Dohlen}, {Sauvage},
  {Milli}, {Zins}, {Petit}, {Wahhaj}, {Cantalloube}, {Caillat}, {Costille}, {Le
  Merrer}, {Carlotti}, {Beuzit}, \& {Mouillet}}]{Vigan2019}
{Vigan}, A., {N'Diaye}, M., {Dohlen}, K., {et~al.} 2019, Astron.\ Astrophys.,
  629, A11

\bibitem[{{Wahhaj} {et~al.}(2021){Wahhaj}, {Milli}, {Romero}, {Cieza}, {Zurlo},
  {Vigan}, {Pe{\~n}a}, {Valdes}, {Cantalloube}, {Girard}, \&
  {Pantoja}}]{Wahhaj2021}
{Wahhaj}, Z., {Milli}, J., {Romero}, C., {et~al.} 2021, \aap, 648, A26

\end{thebibliography}

%-----------------------------------------------------------------

\end{document}